\documentstyle[aps,preprint,epsfig,graphics]{revtex} 
\begin{document} 
\title{Critical temperature of Bose--Einstein condensation\\
in trapped atomic Bose--Fermi mixtures} 
\author{A. P. Albus$^{1}$, S. Giorgini$^{2}$, F. Illuminati$^{3}$, 
and L. Viverit$^{2,4}$} 
\address{\mbox{}$^{1}$Institut f\"{u}r Physik, Universit\"{a}t Potsdam, 
D--14469 Potsdam, Germany\\
\mbox{}$^{2}$Dipartimento di Fisica, Universit\`{a} di Trento, 
and Istituto Nazionale per la Fisica della Materia, I--38050 Povo (TN), 
Italy\\
\mbox{}$^{3}$Dipartimento di Fisica, Universit\`{a} di Salerno, 
and Istituto Nazionale per la Fisica della Materia, I--84081 Baronissi
(SA), Italy\\
 \mbox{}$^{4}$Dipartimento di Fisica, Universit\`{a} di Milano,
via Celoria 16, I--20133 Milano, Italy}
\date{\today} 
\maketitle 
 
\begin{abstract} 
{\it We calculate the shift in the critical temperature of 
Bose-Einstein condensation for a dilute Bose-Fermi mixture  
confined by a harmonic potential to lowest order in both  
the Bose-Bose and Bose-Fermi coupling constants. The relative  
importance of the effect on the critical temperature of the  
boson-boson and boson-fermion interactions is investigated  
as a function of the parameters of the mixture. The possible  
relevance of the shift of the transition temperature in  
current experiments on trapped Bose-Fermi mixtures is
discussed.} 
\end{abstract} 
 
\vspace{0.4cm} 
 
\narrowtext 
 
\vspace{1.2cm} 
 
\section{Introduction} 
 
The achievement of Bose-Einstein condensation in ultracold, trapped 
dilute alkali gases, beyond realizing a striking and spectacular 
experimental confirmation of a long-standing, fundamental prediction  
of quantum theory, has initiated and stimulated as well a whole new 
field of research in the physics of quantum gases in confined geometries 
\cite{RExp,RThe}. Nowaday, many different atomic species and isotopes  
can be succesfully cooled and trapped in the gaseous state, to 
investigate  
in exceptional conditions of purity and isolation the properties of  
interacting Bose and Fermi systems, or Bose-Fermi mixtures.  
 
In particular, the experimental realizations of 
trapped gaseous mixtures of bosons and fermions are 
both an interesting new instance of a quantum  
many-body system and a very useful 
tool to reach the regime of quantum degeneracy 
for a Fermi gas via sympathetic cooling of the 
fermions by the bosons 
\cite{Hulet,Paris,MIT,Jin,Florence}.
From a theoretical point of view, dilute 
Bose-Fermi mixtures have been the object of recent 
investigations addressing, for example, the determination of 
the density profiles of the two components  
in trapped systems \cite{TBF}, the problem of 
stability and phase separation  
\cite{VIVPS}, and the effect of boson-fermion 
interactions on the dynamics \cite{YIP} and on 
the ground-state properties \cite{US} of the mixture. 
 
Boson-fermion interactions in a Bose-Fermi mixture 
can induce a net attractive  
interaction between the fermions, 
thus introducing a further mechanism toward  
the achievement of the BCS transition in trapped 
Fermi gases \cite{II}. In this  
paper we address the reverse problem, i.e. 
how the transition temperature 
of Bose-Einstein condensation is 
affected by the presence of the fermions  
in a trapped mixture. 
The shift of the transition temperature $T_c$ due to  
interactions in a pure trapped Bose 
system has been calculated within the  
mean-field approximation in Ref. \cite{GioPit96}. 
In the present work we extend  
the perturbative methods of Ref. \cite{GioPit96} 
to obtain the shift of $T_c$  
to lowest order in both the 
Bose-Bose and Bose-Fermi coupling constants. 
To this order the effect on $T_c$ of boson-boson and boson-fermion 
interactions are independent and add linearly. 
The relative importance of the two effects  
depends on the relevant parameters of the trapped mixture: 
number of bosons and  
fermions in the trap, ratio of the masses and of the oscillator 
frequencies for 
the two species and the ratio of the Bose-Bose and Bose-Fermi 
coupling constants.    
The calculation is carried out in local density approximation 
which is valid provided that 
the number of bosons and fermions in the trap is large. 
Finite size effects have not been included in the
present treatment.
   
The plan of the paper is as follows: in Section II we generalize 
the scheme derived for pure Bose systems in
Ref. \cite{GioPit96}, to include the effects of boson-fermion  
interactions. In Section III we derive analytical results 
for the shift of $T_c$ in  
the limits of a highly degenerate (Thomas-Fermi) and a 
classical (Boltzmann) Fermi gas,
and we provide the full numerical solution 
for the intermediate regimes. 
We finally compare the theoretical predictions with the current 
experimental situations and we draw our conclusions.
 
\section{Theory} 
 
In a non-interacting Bose gas confined by 
the external harmonic potential 
$V_{ext}^B({\bf r})=m_B(\omega_x^2x^2
+\omega_y^2y^2+\omega_z^2z^2)/2$,  
the critical temperature for Bose-Einstein 
condensation (BEC) is given by 
\begin{equation} 
k_BT_c^0 = \hbar\omega_B\;  
\left(\frac{N_B}{\zeta(3)}\right)^{1/3} 
\simeq 0.94 \hbar\omega_B\;N_B^{1/3}\;, 
\label{Tc0} 
\end{equation} 
where $\omega_B=(\omega_x\omega_y\omega_z)^{1/3}$ is 
the geometric mean of the  
oscillator frequencies and $m_B$, $N_B$ are respectively 
the particle mass and  
the number of bosons in the trap. The above result is obtained using   
the local density approximation (LDA), where the temperature 
of the gas is assumed  
to be much larger than the spacing between single 
particle levels: $k_BT\gg  
\hbar\omega_{x,y,z}$. In this case the density of thermal 
atoms can be written as 
\begin{equation} 
n_B^0({\bf r}) = (\lambda_T^B)^{-3} g_{3/2}(\exp\{-[V_{ext}^B({\bf r})
- \mu_B]/k_BT\}) \;,  
\label{nBzero} 
\end{equation} 
where $\lambda_T^B=\hbar(2\pi/m_Bk_BT)^{1/2}$ is the 
boson thermal wavelength, and   
$g_{3/2}(x)=\sum_{n=1}^{\infty} x^n/n^{3/2}$ is the standard 
Bose function of order $3/2$.  
At $T=T_c^0$ the boson chemical potential takes the critical 
value $\mu_B=\mu_c^0=0$,  
corresponding to the bottom of the external potential, 
and the density $n_B^0(0)$ in  
the center of the trap satisfies the critical 
condition $n_B^0(0)(\lambda_{T_c^0}^B)^3 
=\zeta(3/2)\simeq 2.61$ holding for a homogeneous system.   
 
Finite size effects modify the prediction of the critical 
temperature (\ref{Tc0})  
resulting in a reduction of $T_c^0$. 
The first correction due to the finite number  
of atoms in the trap is given by \cite{Grossmann-Holthaus}:
\begin{equation} 
\left( \frac{\delta T_c}{T_c^0} \right)_{fs} =
-\frac{\zeta(2)}{2\zeta(3)^{2/3}}
\frac{\bar{\omega}_B} 
{\omega_B}N_B^{-1/3}\simeq - 0.73 
\frac{\bar{\omega}_B}{\omega_B} N_B^{-1/3} \;, 
\label{delTc0} 
\end{equation} 
where $\bar{\omega}_B=(\omega_x+\omega_y+\omega_z)/3$ is 
the arithmetic mean of the oscillator frequencies. 
 
Interparticle interactions have an effect on the BEC 
transition temperature as well. 
The presence of repulsive interactions 
has the effect of expanding the 
atomic cloud, with a consequent decrease of the density. 
Lowering the peak density has 
then the effect of lowering the critical temperature. 
On the contrary, attractive 
interactions produce an increase of the density and thus an 
increase of $T_c$. This  
effect, which is absent in the case of a uniform gas where 
the density is kept fixed, can be easily estimated within 
mean-field theory. For pure bosonic systems  
the shift $\delta T_c=T_c-T_c^0$ has been calculated 
in Ref. \cite{GioPit96}, 
\begin{equation} 
\left( \frac {\delta T_c}{T_c^0} \right)_{BB} 
= - 1.33 \;\frac{a_{BB}}{\ell_B}\;N_B^{1/6} \;, 
\label{deltaTc1} 
\end{equation} 
to lowest order in the coupling constant 
$g_{BB}=4\pi\hbar^2a_{BB}/m_B$. 
In the above equation $a_{BB}$ is the boson-boson $s$-wave 
scattering length and $\ell_B=\sqrt{\hbar/m_B\omega_B}$ is the 
harmonic oscillator length. Result  
(\ref{deltaTc1}) has been obtained within LDA 
and neglects finite size effects. 
 
In the case of trapped Bose-Fermi mixtures the shift 
of $T_c$, due to both Bose-Bose  
and Bose-Fermi couplings can be calculated in mean-field 
approximation using the  
methods of Ref. \cite{GioPit96}. 
The transition temperature $T_c$ of a trapped Bose  
gas is defined by the normalization condition  
\begin{equation} 
N_B=\int d{\bf r} \; n_B({\bf r},T_c,\mu_c) \;, 
\label{Tc} 
\end{equation} 
where $n_B$ is the thermal density of bosons 
and $\mu_c$ is the critical value of the boson 
chemical potential. 
Within LDA the boson density above $T_c$ is given by 
\begin{equation} 
n_B({\bf r}) = 
(\lambda_T^B)^{-3} g_{3/2}(\exp\{-[V_{eff}^B({\bf r})
-\mu_B]/k_BT\}) \;,  
\label{nB} 
\end{equation} 
where 
\begin{equation} 
V_{eff}^B({\bf r})=V_{ext}^B({\bf r})
+ 2g_{BB}n_B({\bf r})+g_{BF}n_F({\bf r}) \;, 
\label{Veff} 
\end{equation} 
is the effective potential acting on the bosons 
which is generated by the external field  
$V_{ext}^B$ and by the mean field produced by interactions 
with the other bosons  
and with the fermions. Notice the factor 2 present in the 
Bose-Bose contribution and absent in  
the Bose-Fermi term due to exchange effects. 
In the above equation $n_F({\bf r})$  
is the fermion density and $g_{BF}=2\pi\hbar^2a_{BF}/m_R$ 
is the Bose-Fermi coupling 
constant, fixed by the boson-fermion $s$-wave 
scattering length $a_{BF}$ and by the  
reduced mass $m_R=m_B m_F/(m_B+m_F)$, 
where $m_F$ is the fermion mass. 
 
For a fixed value of the boson chemical potential 
$\mu_B$ and a fixed temperature $T$, the boson density  
(\ref{nB}) can be expanded to first order 
in $g_{BB}$ and $g_{BF}$ as 
\begin{equation} 
n_B({\bf r},T,\mu_B)=n_B^0({\bf r},T,\mu_B)
- [2g_{BB}n_B^0({\bf r})+g_{BF}n_F^0({\bf r})] 
\frac{\partial n_B^0}{\partial\mu_B} \;, 
\label{nB1} 
\end{equation}     
in terms of the non-interacting boson (\ref{nBzero}) 
and fermion density 
\begin{equation} 
n_F^0({\bf r}) = 
(\lambda_T^F)^{-3} f_{3/2}(\exp\{-[V_{ext}^F({\bf r})
-\mu_F]/k_BT\}) \;.  
\label{nFzero} 
\end{equation} 
In the above equation $\lambda_T^F=\hbar(2\pi/m_Fk_BT)^{1/2}$ 
is the fermion thermal wavelength, and $f_{3/2}(x)$ is the Fermi 
function of order $3/2$ defined as 
\begin{equation} 
f_{3/2}(x)=\frac{2}{\sqrt{\pi}}\int_0^\infty  
dz\frac{\sqrt{z}}{e^{z}/x+1} \;. 
\label{Fermifun} 
\end{equation} 
Result (\ref{nFzero}) has been obtained in 
LDA for a Fermi gas in the trapping potential  
$V_{ext}^F({\bf r}) = 
m_F(\omega_x^{\prime\, 2}x^2+\omega_y^{\prime\, 2}y^2
+ \omega_z^{\prime\, 2} z^2)/2$. The fermion
chemical potential 
$\mu_F$ is fixed by the normalization condition 
\begin{equation} 
N_F=\int d{\bf r} \; n_F^0({\bf r}) \;, 
\label{Ferminorm} 
\end{equation} 
where $N_F$ is the total number of fermions in the trap. 
The condition of validity for LDA 
requires the Fermi temperature of the fermionic system to 
be much larger than the harmonic 
oscillator energies $k_BT_F\gg \hbar\omega_{x,y,z}^\prime$. 
For a non-interacting trapped Fermi 
system the Fermi temperature, 
or equivalently the Fermi energy, 
is given by $k_BT_F=\epsilon_F
=\hbar\omega_F(6N_F)^{1/3}$, 
where $\omega_F =
(\omega_x^\prime\omega_y^\prime\omega_z^\prime)^{1/3}$
is the geometric mean of the fermion oscillator frequencies.
 
To first order in $g_{BB}$ and $g_{BF}$, the critical value 
$\mu_{c}$ of the boson chemical potential can be written as 
\begin{equation} 
\mu_c=\mu_c^0+2g_{BB}n_B^0({\bf r}=0)+g_{BF}n_F^0({\bf r}=0) \;. 
\label{muc} 
\end{equation}   
By writing $T_c=T_c^0+\delta T_c$, one can 
expand Eq. (\ref{Tc}) obtaining the following result 
for the total relative shift of the condensation temperature:

\begin{eqnarray} 
\frac{\delta T_c}{T_c^0} = 
\left(\frac{\delta T_c}{T_c^0}\right)_{BB} + 
\, \left(\frac{\delta T_c}{T_c^0}\right)_{BF} =
&-& \frac{2g_{BB}}{T_c^0}\frac{\int d{\bf r}\; 
\partial n_B^0/\partial \mu_B 
[n_B^0({\bf r}=0)-n_B^0({\bf r})]}{\int d{\bf r}\; 
\partial n_B^0/\partial T}  
\nonumber \\ 
&-& \frac{g_{BF}}{T_c^0}\frac{\int d{\bf r}\; 
\partial n_B^0/\partial \mu_B 
[n_F^0({\bf r}=0)-n_F^0({\bf r})]}{\int d{\bf r}\; 
\partial n_B^0/\partial T} \;, 
\label{deltaTc2} 
\end{eqnarray}

\noindent where the derivatives of the non-interacting 
boson and fermion densities $n_B^0$ and $n_F^0$ are 
evaluated at the ideal critical point $\mu_c^0 = 0$, 
$T=T_c^0$. 
The first term $\left(\delta T_c/T_c^0\right)_{BB}$ 
in the above equation accounts for interaction 
effects among the bosons and 
coincides with the shift (\ref{deltaTc1}). 
The second term $\left(\delta 
T_c/T_c^0\right)_{BF}$ accounts instead for 
interaction effects between bosons and fermions, and
its determination will constitute the  
main result of the present paper. 
Some comments are in order here. (i) The shift  
$\delta T_c$ derived above is a mean-field effect 
which originates from the fact that in a trapped Bose-Fermi  
mixture the total number of bosons and the total number of
fermions are fixed, but not the density profiles 
of the two species.  
This effect is peculiar of trapped systems, since
it vanishes identically in the case of uniform systems, and  
should not be confused with the shift of $T_c$ 
occurring in homogeneous Bose systems, which is instead
due to many-body effects \cite{Baym}. (ii) The shift originating 
from the Bose-Fermi coupling, similarly to the one arising  
from the Bose-Bose one, is negative if $g_{BF}>0$ and is 
positive if $g_{BF}<0$. If $a_{BB}$ and $a_{BF}$  
have opposite sign, the corresponding shifts of $T_c$ go in 
opposite directions. (iii) Result (\ref{deltaTc2})  
holds to lowest order in $g_{BB}$ and $g_{BF}$ and, 
since it has been obtained using LDA, is exact if the  
number of bosons and fermions is large. 
Finite-size corrections are not included in (\ref{deltaTc2}). For a  
finite system, a reasonable estimate of the 
total shift of the critical temperature can be obtained by adding  
to result (\ref{deltaTc2}) the finite-size correction 
(\ref{delTc0}) of the non-interacting model.    

\section{Results} 
 
We now concentrate on the relative 
shift $\left(\delta T_c/T_c^0\right)_{BF}$ due to the boson-fermion 
interaction. First of all we observe that
\begin{equation}
\frac{\partial n_B^0({\bf r})}{\partial \mu_B}=
\frac{1}{(\lambda_{T_c^0}^B)^3k_B T_c^0}\;
g_{1/2}(\exp[-V_{ext}^B({\bf r})/k_BT_c^0]) \;,
\end{equation}
and $T_c^0{\int d{\bf r}\;\partial n_B^0 / \partial T}=3N_B$,
where the derivatives are evaluated at the condensation
point of the non-interacting gas $\mu_c^0 = 0$, $T=T_c^0$. 
Using Eq. (\ref{nFzero}), 
the relative shift can then be rewritten as: 
\begin{eqnarray} 
\nonumber
\left( \frac{\delta T_c}{T_c^0}\right)_{BF} &=&   
- \frac{g_{BF}}{3N_B} 
\frac{1}{(\lambda_{T_c^0}^B)^3(\lambda_{T_c^0}^F)^3k_B T_c^0}\\
\nonumber
&\times&\int d{\bf r}\; g_{1/2}(\exp[-V_{ext}^B({\bf r})/k_BT_c^0])\\
&\times&\Large[ f_{3/2}(\exp\{\mu_F/k_BT_c\}) - 
f_{3/2}(\exp\{[\mu_F-V_{ext}^F({\bf r})]/k_BT_c^0\})\Large] \;.
\label{BFshift} 
\end{eqnarray} 

In the following we shall assume that even if the trapping potentials 
of bosons and fermions can have different oscillator frequencies,
nevertheless $\omega_x/\omega_x'=\omega_y/\omega_y'
=\omega_z/\omega_z'=\omega_B/\omega_F$, i.e. the anisotropy is 
the same for the bosonic and fermionic trapping potentials.
This is always the case in today's experiments, and assuming
otherwise would introduce unnecessary complications.
In fact, the assumption of equal anisotropies
holds in general in magnetic traps since the confining potentials
depend only on the (common) external magnetic field, 
the magnetic moments, and the masses of the atoms.
Eq. (\ref{BFshift}) contains the fermion chemical potential 
$\mu_F(N_F,T_c^0)$ which has to be determined 
from Eq. (\ref{Ferminorm}). Eqs. (\ref{BFshift})
and (\ref{Ferminorm}) have then to be solved 
simultaneously. We notice that Eq. (\ref{Ferminorm}) 
can be rewritten in dimensionless form as
\begin{eqnarray} 
\tilde{T}_F^3
= 3\int_0^{\infty}\;dt\;\frac{t^2}{\exp(t-\tilde{\mu}_F)+1} \;,
\label{Fermino2} 
\end{eqnarray} 
where we have introduced the reduced chemical 
potential $\tilde{\mu}_F=\mu_F/k_BT_c^0$ 
and the reduced Fermi temperature $\tilde{T}_F=T_F/T_c^0$. 
Eq. (\ref{Fermino2}) reveals that 
$\tilde{\mu}_F$ is only a function of $\tilde{T}_F$, 
which in turn is a measure of 
the degeneracy of the Fermi gas at $T=T_c^0$. 
In terms of $\tilde{\mu}_F$ and $\tilde{T}_F$ 
Eq. (\ref{BFshift}) then becomes 
\begin{eqnarray} 
\nonumber
\left( \frac{\delta T_c}{T_c^0} \right)_{BF} &=&   
- \frac{4\pi g_{BF}}{3N_B} 
\frac{R_F^3}{(\lambda_{T_c^0}^B)^3(\lambda_{T_c^0}^F)^3k_B T_c^0}\\
\nonumber
&\times&\int\;ds\; s^2\,g_{1/2}(\exp\{-\tilde{T}_F\,\alpha\,s^2\})\\
&\times& \Large[ f_{3/2}(\exp\{\tilde{\mu}_F\})
- f_{3/2}(\exp\{\tilde{\mu}_F-\tilde{T}_F\,s^2\}) \Large] \;.
\label{BFshift0} 
\end{eqnarray} 
In writing Eq. (\ref{BFshift0}) we have rescaled 
each integration coordinate
by the appropriate Thomas-Fermi radius of the fermion cloud
$R_i'=(2\epsilon_F/m_F\omega_i'^{\,2})^{1/2}$. We have then
introduced the mean Fermi radius $R_F=(R_x'R_y'R_z')^{1/3}$
and named $\alpha=m_B\omega_B^2/m_F\omega_F^2$. 
Since $\tilde{\mu}_F$ depends only
on $\tilde{T}_F$ through Eq. (\ref{Fermino2}), 
the integral in Eq. (\ref{BFshift0}) above  
depends only on the values of the two 
parameters $\tilde{T}_F$ and $\alpha$.

The system of Eqs. (\ref{BFshift0}) and  (\ref{Fermino2}) for 
general $\tilde{T}_F$ and $\alpha$ can only be solved numerically,
and later we shall present the full numerical
results for some specific choices of the parameters. However,
analytical solutions exist in two limits: when $\tilde{T}_F\gg 1$
(i.e. $T_F\gg T_c^0$)
where the Fermi gas is completely degenerate at $T=T_c^0$ 
(Thomas-Fermi regime), and when
$\tilde{T}_F\ll 1$ (i.e. $T_F\ll T_c^0$)
so that at $T_c^0$ fermions behave as a classical gas
(Boltzmann regime).

In order to clarify the connection between the two limits
and the general numerical solution, it is useful to further
manipulate Eq. (\ref{BFshift0}). By explicitly evaluating the 
prefactor, it can be finally recast in the convenient form 
\begin{equation} 
\left( \frac{\delta T_c}{T_c^0} \right)_{BF} =   
- \frac{2^{5/3}}{3^{5/6}\pi\zeta(3)}\left(\frac{m_F}{m_B}+1\right)\;
\frac{a_{BF}}{\ell_F}N_F^{1/6}\;\cdot F(\tilde{T}_F,\alpha) \;,
\label{BFshift01} 
\end{equation} 
where
\begin{eqnarray} 
\nonumber
F(\tilde{T}_F,\alpha) &=&   
\alpha^{3/2}\tilde{T}_F\,\int\;ds\; s^2\,
g_{1/2}(\exp\{-\tilde{T}_F\,\alpha\,s^2\})\\
&\times& \Large[ f_{3/2}(\exp\{\tilde{\mu}_F\})
- f_{3/2}(\exp\{\tilde{\mu}_F-\tilde{T}_F\,s^2\}) \Large] \;,
\label{functionF} 
\end{eqnarray} 
and $\ell_F=\sqrt{\hbar/m_F\omega_F}$ is the fermionic oscillator 
length. Notice the formal analogy between Eq. (\ref{BFshift01}) 
and Eq. (\ref{deltaTc1}) for the shift $(\delta T_c/T_c^0)_{BB}$ 
due to the boson-boson interactions alone.

Let us begin by considering the
Thomas-Fermi limit ($\tilde{T}_F\gg 1$). In this limit the chemical 
potential of the fermions $\mu_F$ tends to the Fermi energy 
$\epsilon_F=k_B T_F$. 
Thus $\tilde{\mu}_F\simeq \tilde{T}_F\gg 1$. The limit of the
Fermi functions in Eq. (\ref{Fermifun}) for $x\to \infty$ is
$f_{3/2}(x)\approx 4(\ln x)^{3/2}/3\sqrt{\pi}$.
This implies that the density profile of the fermion cloud takes the 
well known Thomas-Fermi shape
\begin{equation} 
n_F^0({\bf r})=n_F^0(0)\,
\left[ 1-(x/R'_x)^2-(y/R'_y)^2-(z/R'_z)^2 \right] ^{3/2} \;,
\end{equation} 
with $n_F^0(0)=(2\epsilon_Fm_F/\hbar^2)^{3/2}/(6\pi^2)$, whenever the 
expression inside the square brackets is positive, 
and $n_F^0({\bf r})=0$ otherwise.

The function $F(\tilde{T}_F,\alpha)$ then goes to the 
limiting form
\begin{eqnarray} 
\nonumber
F(\tilde{T}_F,\alpha)&\to&  \frac{4}{3\sqrt{\pi}}
\;\alpha^{3/2}(\tilde{T}_F)^{5/2}
\sum_{n=1}^{\infty}\frac{1}{n^{1/2}}\int_0^1\;ds\,s^2\\
&\times& e^{-n\,\tilde{T}_F\alpha\,s^2}
[1-(1-s^2)^{3/2}] \;,
\label{BFshift2} 
\end{eqnarray} 
since 
$g_{1/2}(x)=\sum_{n=1}^{\infty}x^n/n^{1/2}$. 

We obtained 
Eq. (\ref{BFshift2}) in the limit $\tilde{T}_F\gg 1$. Therefore, 
if $\alpha$ is not too small (so that $\tilde{T}_F\alpha\gg 1$
still holds), then, for every $n$ in the series,
the exponential is non-vanishing only for values of $s\ll 1$,
and we can adopt the expansion $1-(1-s^2)^{3/2}\simeq 3s^2/2$. 
The integral in Eq. (\ref{BFshift2}) becomes 
$\int_0^1\;ds\,s^4\,e^{-n\,\tilde{T}_F\,\alpha\,s^2}
\simeq 3\sqrt{\pi}/[8(n\,\tilde{T}_F\,\alpha)^{5/2}]$.
Finally, therefore, $F(\tilde{T}_F,\alpha)\to 3\zeta(3)/4\alpha$ 
and the Thomas-Fermi prediction for 
the relative shift reads
\begin{equation} 
\left( \frac{\delta T_c}{T_c^0} \right)_{BF}=
-\frac{3^{1/6}}{2^{1/3}\pi}\left(\frac{m_F}{m_B}+1\right)
\frac{m_F\omega_F^2}
{m_B\omega_B^2}\,\frac{a_{BF}}{\ell_F}\,N_F^{1/6} \;,
\end{equation} 
where $3^{1/6}/(2^{1/3}\pi)\simeq 0.304$. We notice that
in the Thomas-Fermi regime the shift is 
independent of the number of bosons $N_B$ and 
varies as the first inverse power of the parameter 
$\alpha=m_B\omega_B^2/m_F\omega_F^2$.

We now consider the Boltzmann limit for the Fermi gas
($\tilde{T}_F\ll 1$). In this case the chemical potential
$\tilde{\mu}_F$ is large and negative and depends on $\tilde{T}_F$ as:
$\tilde{\mu}_F\approx \ln\{(\tilde{T_F})^3/6\}$.
In the limit $x\to 0$, $f_{3/2}(x)\approx x$, and then
\begin{eqnarray} 
\nonumber
F(\tilde{T}_F,\alpha)&\to& \frac{\alpha^{3/2}(\tilde{T}_F)^4}{6}
\sum_{n=1}^{\infty}\frac{1}{n^{1/2}}\int_0^{\infty}\;ds\;s^2\\
&\times& \left[e^{-n\tilde{T}_F\alpha s^2}
-e^{-(n\tilde{T}_F\alpha+\tilde{T}_F) s^2}\right] \;.
\label{funFbolz}
\end{eqnarray} 

Evaluation of the integrals is straightforward and yields
\begin{equation} 
F(\tilde{T}_F,\alpha) \to \frac{\sqrt{\pi}}{24} (\tilde{T}_F)^{5/2}
\cdot f(\alpha) \;,
\label{funFbolz2}
\end{equation} 
with
\begin{equation} 
f(\alpha) =
\sum_{n=1}^{\infty}\left(\frac{1}{n^{2}}
-\frac{1}{n^{1/2}(n+\alpha^{-1})^{3/2}}\right) \;,
\label{fofalpha} 
\end{equation} 
so that:
\begin{equation} 
\left( \frac{\delta T_c}{T_c^0} \right)_{BF} =
-\frac{1}{2^{4/3}\pi^{1/2}3^{11/6}\zeta(3)}\left(\frac{m_F}{m_B}+1\right)
\frac{a_{BF}}{\ell_F}\,N_F^{1/6}(\tilde{T}_F)^{5/2}\cdot f(\alpha) \;,
\label{Boltzmannshift}
\end{equation} 
where the numerical prefactor is $\simeq 0.025$.

We notice that $f(\alpha)$ in Eq. (\ref{fofalpha}) is a monotonically
decreasing function of $\alpha$. 
In particular, one finds the following behaviours:
$f(\alpha\to 0)= \pi^2/6$, $f(1)\simeq 0.85$,
and $f(\alpha\to \infty)=  3\zeta(3)/2\alpha$.
As one should expect, in the Boltzmann limit ($\tilde{T}_F \ll 1$), 
the Bose-Fermi shift is negligible.

\begin{figure}[h]
\begin{center}
\epsfxsize=8.5cm
%\rotatebox{-90}{\includegraphics*[height=9cm,width=13cm]{Figure1.eps}}
\rotatebox{-90}{\epsfbox{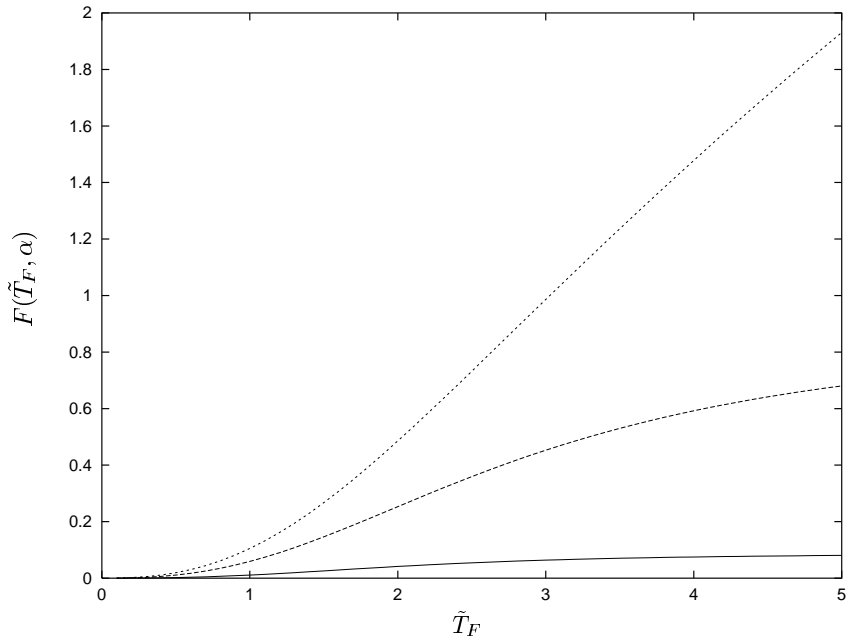}}
\end{center}
\caption{Dimensionless function $F(\tilde{T}_F,\alpha)$
as a function of $\tilde{T}_F$ for
the values $\alpha=0.1$ (dotted line), $\alpha=1.0$ (dashed line), and
$\alpha=10$ (solid line).}
\label{MedTC}
\end{figure}

We now turn to the full numerical solution of
Eqs. (\ref{BFshift01}), (\ref{functionF}), 
and (\ref{Fermino2}) for more general values of 
the degeneracy parameter $\tilde{T}_F$.
In Fig. \ref{MedTC} we show the dimensionless function 
$F(\tilde{T}_F,\alpha)$ as a function of 
$\tilde{T}_F$ for three different values of 
the parameter $\alpha=m_B\omega_B^2/m_F\omega_F^2$,
$\alpha=0.1$, $1$, and $10$.
For fixed $\alpha$, $F(\tilde{T}_F,\alpha)$ is a monotonically 
nondecreasing function of $\tilde{T}_F$, which saturates for 
$\tilde{T}_F\to \infty$ at the value predicted in the Thomas-Fermi
regime $3\zeta(3)/4\alpha\simeq 0.9\;\alpha^{-1}$. 
For fixed $\tilde{T}_F$,
$F(\tilde{T}_F,\alpha)$ increases by decreasing $\alpha$. 
For the largest value of $\alpha$ 
($\alpha=10$) the function $F$ reaches 
its asymptotic Thomas-Fermi value
already at $\tilde{T}_F\simeq 5$. 
For $\alpha=1$ and $\alpha=0.1$ the 
function saturates for larger 
values of $\tilde{T}_F$ not shown in the 
figure. The reason for this difference 
can be understood by recalling that
the Thomas-Fermi result requires not only $\tilde{T}_F\gg 1$, but
also $\tilde{T}_F\gg \alpha^{-1}$ (see the discussion below Eq. 
(\ref{BFshift2})).

The physically relevant regimes in current experiments fall
roughly around $\alpha \simeq 1$ and $\tilde{T}_F \simeq 1$.
In this respect, a particularly interesting situation
is the one realized in the Florence experiment \cite{Florence}, 
where a quantum
degenerate trapped atomic mixture of fermionic ${}^{40}$K and
bosonic ${}^{87}$Rb has been recently produced. One of the
appealing features of this system is that the measured
boson-fermion scattering length is large and negative: 
$a_{BF}=-22$ nm, giving rise to a fairly strong
attractive boson-fermion interaction. The shift 
$\left(\delta T_{c}/T_{c}^{0} \right)_{BF}$ is thus
positive and opposite to the shift 
$\left( \delta T_{c}/T_{c}^{0} \right)_{BB}$, since for
pure ${}^{87}$Rb the boson-boson scattering length is
$a_{BB}=6$ nm, giving rise to a repulsive boson-boson
interaction. In the Florence experiment the two atomic
species are magnetically trapped, and are both prepared
in their doubly polarized spin state. These states 
experience the same trapping potential so that 
$\alpha=m_B\omega_B^2/m_F\omega_F^2=1$, while the
number of bosons and of fermions are respectively
$N_{B} = 2 \times 10^{4}$, $N_{F} = 10^{4}$, so that
$N_{F}/N_{B} = 0.5$, and $\tilde{T}_{F} = T_{F}/T_{c}^{0}
\simeq 2.3$. For the conditions of the Florence experiment 
the shift (\ref{deltaTc1}) due to the boson-boson coupling turns 
out to be: $\left( \delta T_c/T_c^0 \right)_{BB}\simeq -0.037$, 
and is comparable with the shift (\ref{delTc0}) due to finite 
size effects,
which is given by: $\left( \delta T_{c}/T_c^0
\right)_{fs} = -0.044$. 
For $\alpha=1$ at 
$\tilde{T}_F\simeq 2.3$ the function $F$ is at about 1/3 of its
asymptotic value in the Thomas-Fermi regime, resulting in a 
Bose-Fermi shift considerably smaller than the Bose-Bose one: 
$\left( \delta T_c/T_c^0 \right)_{BF}\simeq 0.012$. In Fig. 2 
we show the shift  $\left( \delta T_c/T_c^0 \right)_{BF}$ as a 
function of the ratio $N_{F}/N_{B}$, with all the other parameters 
entering Eq. (\ref{BFshift01}) fixed at the values of the Florence 
experiment \cite{Florence}. In the same figure we include as a 
reference value the modulus of the boson-boson relative shift 
$|\left( \delta T_c/T_c^0 \right)_{BB}|$, calculated using the 
values of the parameters given by the Florence experiment.

\begin{figure}[h]
\begin{center}
\epsfxsize=8.5cm
%\rotatebox{-90}{\includegraphics*[height=9cm,width=13cm]{Figure2eps}}
\rotatebox{-90}{\epsfbox{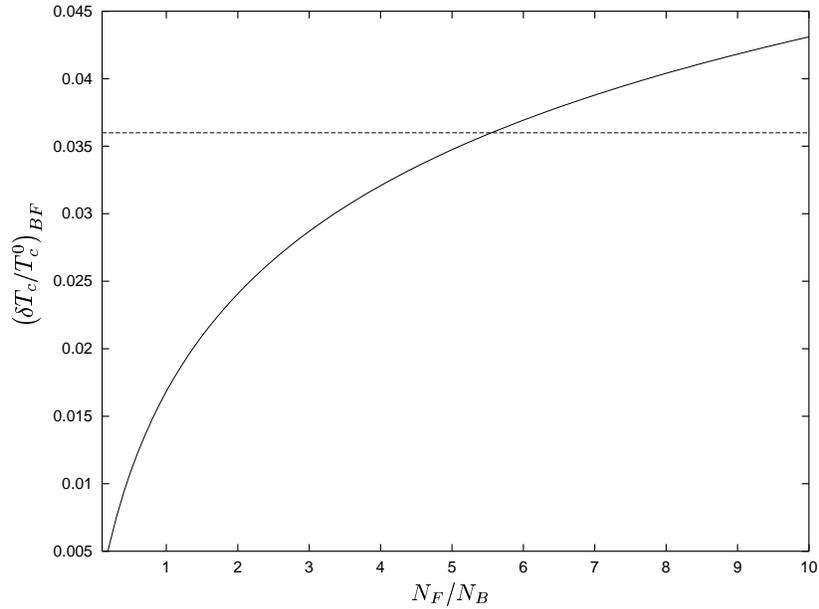}}
\end{center}
\caption{Boson-fermion relative shift $\left( 
\delta T_c/T_c^0 \right)_{BF}$ 
from Eq. (\ref{BFshift01}) (solid line) as a function of 
the ratio $N_{F}/N_{B}$. 
Horizontal dashed line: value of the modulus 
$|\left( \delta T_c/T_c^0 \right)_{BB}|$ of 
the boson-boson shift (\ref{deltaTc1}).
All other parameters, except the number of fermions $N_F$, 
have been fixed at 
the values of the Florence experiment.}
\label{Florenceshift}
\end{figure}

From Fig. \ref{Florenceshift} we see that, while in the present 
experimental situation  
the boson-fermion shift is about $1/3$ of the 
boson-boson one, by increasing the number of trapped 
fermions the two shifts
become comparable at $N_F \simeq 5N_B$. The boson-fermion
shift is instead dominant at still larger values of $N_F$.
It is important to remark that, 
even if the Bose-Fermi shift of the critical temperature 
is a small effect for the present experimental conditions, it
might be observable. Since the fermions can be eliminated from the 
trap, one can look for the differences in the
transition temperature with and without fermions.

In conclusion, we have determined the relative shift of the critical
temperature of Bose-Einstein condensation in a trapped atomic
Bose-Fermi mixture to lowest order in both the boson-boson and
the boson-fermion coupling constants. We have determined numerically
the general behaviour of the boson-fermion shift, and we have provided
full analytical solutions in the quantum degenerate Thomas-Fermi
regime and in the classical Boltzmann regime. We have applied
our predictions to a specific experiment
(the Florence experiment \cite{Florence}, chosen for the 
interesting value of the Bose-Fermi scattering length), 
and discussed the relative importance of 
the shifts due to boson-boson and boson-fermion interactions.


\begin{references} 
 
\bibitem{RExp} W. Ketterle, D. S. Durfee, and D. M. Stamper-Kurn, in  
{\it Bose-Einstein Condensation in Atomic Gases}, Proceedings of the  
International School of Physics ``Enrico Fermi'' edited by M. Inguscio,  
S. Stringari, and C. E. Wieman (SIF, Bologna, 1999); E. A. Cornell, 
J. R. Ensher, and C. E. Wieman, {\it ibid}. 
 
\bibitem{RThe} F. Dalfovo, S. Giorgini, L. P. Pitaevskii, 
and S. Stringari, Rev. Mod. Phys. {\bf 71}, 463 (1999).   
 
\bibitem{Hulet} A. G. Truscott, K. E. Strecker, W. I.
McAlexander, G. B. Partridge, and R. G. Hulet,
Science {\bf 291}, 2570 (2001).

\bibitem{Paris} F. Schreck, L. Khaykovich, K.L. Corwin, 
G. Ferrari, T. Bourdel, J. Cubizolles, and C. Salomon, 
Phys. Rev. Lett. {\bf 87}, 080403 (2001).
 
\bibitem{MIT} Z. Hadzibabic, C.A. Stan, K. Dieckmann, 
S. Gupta, M. W. Zwierlein, A. G\"orlitz, and W. Ketterle,  
Phys. Rev. Lett. {\bf 88}, 160401 (2002).
 
\bibitem{Jin} J. Goldwin, S. B. Papp, B. DeMarco, and
D. S. Jin, Phys. Rev. A {\bf 65}, 021402(R) (2002).

\bibitem{Florence} G. Roati, F. Riboli, G. Modugno, 
and M. Inguscio, Phys. Rev. Lett. {\bf 89}, 150403 (2002). 
 
\bibitem{TBF} K. M\o lmer, Phys. Rev. Lett. {\bf 80}, 1804 (1998); 
M. Amoruso, A. Minguzzi, S. Stringari, M. P. Tosi, and L. Vichi, 
Eur. Phys. J. D {\bf 4}, 261 (1998); N. Nygaard and  
K. M\o lmer, Phys. Rev. A {\bf 59}, 2974 (1999);
T. Miyakawa, K. Oda, T. Suzuki, and H. Yabu, J. Phys. Soc.
Jpn. {\bf 69}, 2779 (2000).
 
\bibitem{VIVPS} L. Viverit, C. J. Pethick, and H. Smith, 
Phys. Rev. A {\bf 61}, 053605 (2000);  
X. X. Yi and C. P. Sun, Phys. Rev. A {\bf 64}, 043608 (2001); 
R. Roth and H. Feldmeier, 
Phys. Rev. A {\bf 65}, 021603(R) (2002); 
R. Roth, Phys. Rev. A {\bf 66}, 013614 (2002);
Z. Akdeniz, A. Minguzzi, P. Vignolo, and M. P. Tosi,
Phys. Rev. A {\bf 66}, 013620 (2002).
 
\bibitem{YIP} S. K. Yip, Phys. Rev. A {\bf 64}, 023609 (2001);
H. Pu, W. Zhang, M. Wilkens, and P. Meystre, Phys. Rev. Lett.
{\bf 88}, 070408 (2002); T. Sogo, T. Miyakawa, T. Suzuki,
and H. Yabu, Phys. Rev. A {\bf 66}, 013618 (2002).
 
\bibitem{US} A. P. Albus, S. A. Gardiner, 
F. Illuminati, and M. Wilkens, Phys. Rev. A {\bf 65},  
053607 (2002); L. Viverit and S. Giorgini, 
preprint cond-mat/0207260 (2002). 
 
\bibitem{II} M. J. Bijlsma, B. A. Heringa, and H. T. C. Stoof, 
Phys. Rev. A {\bf 61}, 053601 (2000);  
H. Heiselberg, C. J. Pethick, H. Smith, and L. Viverit, 
Phys. Rev. Lett. {\bf 85}, 2418 (2000); 
L. Viverit, Phys. Rev. A {\bf 66}, 023605 (2002). 
 
\bibitem{GioPit96} 
S. Giorgini, L. P. Pitaevskii, and S. Stringari, 
Phys. Rev. A {\bf 54}, R4633 (1996). 
 
\bibitem{Grossmann-Holthaus} S. Grossmann and 
M. Holthaus, Phys. Lett. A {\bf 208}, 188 (1995); 
Z. Naturforsch. Teil A {\bf 50}, 323 (1995). 
 
\bibitem{Baym}  
G. Baym, J.-P. Blaizot, M. Holzmann, F. Lalo\"e, and 
D. Vautherin, Eur. Phys. J. B {\bf 24}, 107 (2001);
G. Baym, J. Phys. B {\bf 34}, 4541 (2001), and references
therein.  
 
\end{references}
\end{document}